\documentstyle[aps,epsf]{revtex}

\begin{document}

\begin{flushright}
JLAB-THY-97-21 \\
hep-ph/9706410\\
June 1997 \\
\end{flushright}
\vspace{2cm}

\begin{center}
{\Large \bf Light-Ray Evolution Equations and 
 Leading-Twist Parton Helicity-Dependent
 Nonforward Distributions}
\end{center}
\begin{center}
{I.I.  BALITSKY\footnote{Also at St. Petersburg Institute of Nuclear Physics,
Gatchina, Russia} and 
A.V. RADYUSHKIN\footnote{Also at Laboratory of Theoretical Physics, JINR, Dubna,
Russia}
}  \\
{\em Physics Department, Old Dominion University,}
\\{\em Norfolk, VA 23529, USA}
 \\ {\em and} \\
{\em Jefferson Lab,} \\
 {\em Newport News,VA 23606, USA}
\end{center}
\vspace{2cm}

\begin{abstract}

We discuss  the calculation 
of the evolution kernels $\Delta W_{\zeta}(X,Z)$ 
for the leading-twist nonforward parton distributions
${\cal G}_{\zeta} (X,t)$ sensitive to
parton helicities. We present our results for the  kernels
governing  evolution of  the relevant 
light-ray operators  and describe a simple 
method allowing to 
obtain from them the components of the nonforward
kernels $\Delta W_{\zeta}(X,Z)$.

\end{abstract}

\newpage

\section{Introduction}

Applications of perturbative QCD to
deeply virtual Compton scattering \cite{ji,compton,ji2,npd}
and hard exclusive electroproduction 
processes \cite{bfgms,gluon,cfs,npd}  require a generalization of 
usual parton distributions for the case when
long-distance information
is accumulated in  nonforward matrix
elements $\langle p-r |  {\cal O}(0,z)  |p \rangle |_{z^2=0}$ 
of quark and gluon light-cone  operators.
In refs. \cite{compton,gluon,npd} it was shown 
that such matrix elements can be parametrized by 
two basic  types of nonperturbative functions.
The  double distribution $F(x,y;t)$ 
specifies the light-cone ``plus'' fractions $xp^+$ and $yr^+$ 
of the initial hadron momentum $p$ and the momentum transfer $r$ 
carried by the initial parton. 
Since $r^+$ is proportional to $p^+$: 
$r^+  \equiv \zeta p^+$,  it is possible to introduce 
the  nonforward parton distribution 
${\cal F}_{\zeta}(X;t)$ 
with $X=x+y \zeta$ being the total fraction of the initial hadron momentum 
taken  by the initial  parton\footnote{The off-forward parton distributions introduced
by X. Ji \cite{ji,ji2} (see also \cite{drm}) and 
non-diagonal distributions of Collins, Frankfurt 
and Strikman \cite{cfs} can be  related to 
nonforward distributions (see  \cite{npd}) but 
do not coincide with them.}. 
For processes mentioned above, the parameter $\zeta=1-(p'z)/(pz)$ 
characterizing  the longitudinal 
momentum asymmetry (``skewedness'') of the nonforward 
matrix element takes the values $0 < \zeta < 1$.
 
At  leading twist, there are two light-ray quark  operators
 $\bar \psi (0) \gamma_{\mu} E(0,z;A)  \psi (z)$ and 
 $\bar \psi (0)  \gamma_{\mu} \gamma_5 E(0,z;A)  \psi (z)$,
where    
 $E(0,z;A)$ is the standard path-ordered exponential 
which makes the operators gauge-invariant. 
 In the forward case, 
the first operator is related to  
the  spin-averaged distribution functions $f_a(x)$
while the second one corresponds to   the spin-dependent distribution 
functions $\Delta f_a (x)$. 
The nonforward parton distributions 
related to the $\bar \psi  (0) \gamma_{\mu} E(0,z;A)  \psi (z)$
operators were studied in refs. \cite{compton,gluon,npd}.
In this paper, we
will discuss    flavor-singlet
 parton helicity-dependent  nonforward  distributions 
corresponding  to quark operators
$\bar \psi (0) \gamma_{\mu} \gamma_5 E(0,z;A)  \psi (z)$
and the gluonic operator
$G_{\mu \alpha}(0) E(0,z;A) \tilde G_{\alpha \nu}(z)$ 
mixing with each other  under evolution.

\section{Nonforward distributions}

We  define the  nonforward quark distributions by 
writing  the  relevant  matrix element as 
(cf.\cite{ji,npd})
\begin{eqnarray} 
\hspace{-1cm} 
&&\lefteqn{
\langle \, p'  , s' \,  | \,   \bar \psi_a(0) \hat z \gamma_5
E(0,z;A)  \psi_a(z) 
\, | \,p ,  s \,  \rangle |_{z^2=0}
 } \label{57}  \\ \hspace{-1cm} &&  
= \bar u(p',  s')  \hat z  \gamma_5
 u(p,s) \, \int_0^1   
 \left ( e^{-iX(pz)}
 {\cal G}^a_{\zeta}(X;t)  +  e^{i(X-\zeta)(pz)}
{\cal G}^{\bar a}_{\zeta}(X;t)  \right ) 
 \,  dX \nonumber \\ \hspace{-1cm} &&  
+
\, \frac{(rz)}{M} \,  \bar u(p',s')  \gamma_5
 u(p,s) \, \int_0^1   
  \left ( e^{-iX(pz)}
 {\cal P}^a_{\zeta}(X;t)  +  e^{i(X-\zeta)(pz)}
{\cal P}^{\bar a}_{\zeta}(X;t)  \right ) 
 \,  dX
,
\nonumber
 \end{eqnarray} 
where $t\equiv (p' - p)^2$, $a$ denotes the quark flavor
(here we  consider only the flavor-diagonal
distributions), $M$ is the nucleon mass and $s,s'$ specify the nucleon
 polarization. Throughout the paper, we use the ``hat''
convention $\hat z \equiv z^{\mu} 
\gamma_{\mu}$. 
In Eq.(\ref{57}), we explicitly
separated   quark and  antiquark contributions
(cf. \cite{npd}). 
This definition corresponds to the  parton picture in which 
the initial quark (or antiquark) takes  the momentum $Xp$ from the hadronic
matrix element 
and ``returns'' into it the momentum $(X- \zeta) p$.
Since the fraction $X- \zeta$ is  positive for $X > \zeta$
and negative when  $X < \zeta$,
the nonforward distributions can be divided into two 
components.
In the region $X \geq \zeta$,
one can treat ${\cal G}_{\zeta}^a  (X,t)$ 
as  a generalization of the 
usual  distribution function $\Delta f_a(x)$.
In particular, in the limit  
$t \to 0, \zeta  \to 0$, the  limiting curves 
for ${\cal G}_{ \zeta}(X,t)$ reproduce $\Delta f_a(X)$:
\begin{equation}
 {\cal G}^a_{\zeta=0} \, (X,t=0) = \Delta  f_a(X) \  \ ; \  \  
 {\cal G}^{\bar a}_{\zeta=0} \, (X,t=0) =  \Delta f_{\bar a}(X) 
 . 
\label{60} \end{equation}
On the other hand, in the region  $X < \zeta$, both quarks should be treated
as going out of the nucleon matrix element,
with momenta $Xp$ and $(\zeta - X) p$, respectively.
 Now, one can define $Y=X/\zeta$ and treat 
the function ${\cal G}_{\zeta}^a  (X)$ 
as the distribution amplitude $\Psi_{\zeta}^a(Y)$.
In particular, the ${\cal G}$-part in this region can be written as  
\begin{equation}
\zeta \, \bar u(p')  \hat z 
 u(p) \, \int_0^1   
 \left [ e^{-iY(rz)}
 {\cal G}^a_{\zeta}(\zeta Y)  +  e^{-i(1-Y)(rz)}
{\cal G}^{\bar a}_{\zeta}(\zeta Y)  \right ] 
 \,  dY = \zeta \, \bar u(p')  \hat z 
 u(p) \, \int_0^1   
 e^{-iY(rz)}
 \Psi_{\zeta}^a(Y)
 \,  dY \, , 
\label{61} \end{equation}
where  the distribution amplitude  $\Psi_{\zeta}^a(Y)$ is defined by 
$
\Psi_{\zeta}^a(Y) = {\cal G}_{\zeta}^a (Y\zeta) 
+ {\cal G}_{\zeta}^{\bar a} (\bar Y \zeta)\, .
$
The function  $\Psi_{\zeta}^a(Y)$ gives the probability 
amplitude that  the initial nucleon with momentum $p$ is composed 
of the final nucleon with momentum $p' \equiv p-r$
and a $\bar qq$ pair in which the pair momentum $r$ 
is shared in fractions
$Y$ and $1-Y \equiv \bar Y$.

For gluons, the    nonforward  distribution  ${\cal G}^g_{\zeta}(X;t)$  
is defined  
through the matrix element
\begin{eqnarray} 
\hspace{-1cm} 
&&\lefteqn{
\langle p'  \,  | \,   
z_{\mu}  z_{\nu} G_{\mu \alpha}^a (0) E^{ab}(0,z;A) 
\tilde G_{ \alpha \nu }^b (z)\, | \,p  \rangle |_{z^2=0}
 } \label{63}  \\ \hspace{-1cm} &&  
= \bar u(p')  \hat z \gamma_5
 u(p) \, (z \cdot p) \, \int_0^1   
\frac{i}{2}  \left [ e^{-iX(pz)}
 - e^{i(X-\zeta)(pz)} \right ] 
 {\cal G}^g_{\zeta}(X;t)  
 \,  dX 
\nonumber \\ \hspace{-1cm} &&  
+ \, 
\bar u(p')  \frac {(rz)}{M}\gamma_5
 u(p) (z \cdot p) \, \int_0^1   
\frac{i}{2}  \left [ e^{-iX(pz)}
 - e^{i(X-\zeta)(pz)} \right ] 
 {\cal P}^g_{\zeta}(X;t) 
 \,  dX \,  . 
\nonumber
 \end{eqnarray} 
As usual, $\tilde G_{ \alpha \nu } = \frac12 \epsilon_{\alpha \nu \beta \mu} G^{\beta \mu}$.
Since there are no ``antigluons'',
 the exponentials $e^{-iX(pz)}$ and $ e^{i(X-\zeta)(pz)} $ are  
accompanied here by the same function  ${\cal G}^g_{\zeta}(X;t)$. 
 Again, the contribution from the region $0<X<\zeta$
can be written   as
\begin{equation}
i \bar u(p')  \hat z \gamma_5
 u(p) \, (z \cdot r) \, \int_0^1   
  e^{-iY(rz)} \, \Psi_{\zeta}^g(Y;t)
 \,  dY  + ``{\cal P}" \, {\rm term}, 
\label{64} \end{equation}
with the   $Y \leftrightarrow \bar Y$ antisymmetric generalized
distribution amplitude
$\Psi_{\zeta}^g(Y;t)$ given by
\begin{equation}
\Psi_{\zeta}^g(Y;t) = \frac12 \left ( {\cal   G}^g_{\zeta}(Y \zeta;t)
- {\cal   G}^g_{\zeta}(\bar Y \zeta;t) \right ) \, .
\label{65} \end{equation}
In the formal $t=0$ limit, the nonforward distributions 
${\cal G}^g_{\zeta}(X;t)$, ${\cal P}^g_{\zeta}(X;t)$
convert into the asymmetric distribution functions
${\cal G}^g_{\zeta}(X)$, ${\cal P}^g_{\zeta}(X)$.
Finally, in the $\zeta =0$ limit, ${\cal G}^g_{\zeta}(X)$
reduces to the usual polarized gluon density
\begin{equation}
{\cal G}^g_{\zeta=0}(X) = X \Delta g(X).
\label{66} \end{equation}
Under pQCD evolution, 
 the gluonic  operator
\begin{equation}
{\cal O}_{ g} (uz,vz) =
z_{\mu}  z_{\nu} 
G_{\mu \alpha}^a (uz) E^{ab}(uz,vz;A) 
\tilde G_{\alpha \nu}^b (vz) 
 \label{73} \end{equation}
 mixes  with the flavor-singlet 
quark operator
\begin{equation}
{\cal O}_{ Q}(uz,vz) =   \sum_{a=1}^{N_f} {\cal O}_a^{(+)}(uz,vz) 
 \label{67} \end{equation}
where 
\begin{equation}
{\cal O}_a^{(+)}(uz,vz)  =  \frac{1}{2}
\biggl [ \bar \psi_a(uz) 
\hat z \gamma_5 E(uz,vz;A)  \psi_a(vz)
+ \bar \psi_a(vz)  \hat z \gamma_5 E(vz,uz;A) \psi_a(uz) \biggr ] \, . 
 \label{68} \end{equation}
The nonforward  distribution function
${\cal G}_{\zeta}^Q (X;t)$ for the 
flavor-singlet quark combination (\ref{67})
\begin{equation} 
\langle \,  p',s' | \, {\cal O}_{ Q}(uz,vz)\, | \, p,s \rangle |_{z^2=0}  = 
 \bar u(p',s')  \hat z \gamma_5
 u(p,s)   \int_0^1    \, 
 \frac{1}{2}  \left [ e^{-ivX(pz)+iuX'(pz)} + e^{ivX'(pz) 
- iuX(pz)}\right ]
{\cal G}_{\zeta}^Q (X;t)
 \,  dX \, + ``{\cal P}" \, {\rm term},
\label{69} \end{equation}
(where $X' \equiv X - \zeta$) 
can be expressed as the  sum of ``$a+\bar a$'' 
distributions:
\begin{equation}
 {\cal G}_{\zeta}^Q (X;t) =   \sum_{a=1}^{N_f}
({\cal G}_{\zeta}^a (X;t) +
{\cal G}_{\zeta}^{\bar a} (X;t)) \, . 
\label{70} \end{equation}
Writing the  contribution from the $0<X<\zeta$ region as
\begin{equation}
\zeta \bar u(p')  \hat z  \gamma_5
 u(p) \, (z \cdot r) \, \int_0^1   
  e^{-iY(rz)} \Psi_{\zeta}^Q(Y;t)
 \,  dY  + ``{\cal P}" \, {\rm term}, 
\label{64A} \end{equation}
we introduce the flavor-singlet quark
distribution amplitude $\Psi_{\zeta}^Q(Y;t)$ 
which has the  symmetry property 
$\Psi_{\zeta}^Q(Y;t) = \Psi_{\zeta}^Q(\bar Y;t)$  with respect to
the $Y \leftrightarrow \bar Y$ transformation.

\section{ Evolution equations for light-ray operators}

Near  the light cone $z^2 \sim 0$, the bilocal operators  ${\cal O}(uz,vz)$
develop logarithmic singularities $\ln z^2$.  Calculationally, these
singularities manifest themselves as  ultraviolet divergences for operators
taken on the light cone. 
The divergences are removed by  a 
subtraction  prescription characterized 
by some  scale $\mu$: 
${\cal G}_{\zeta} (X;t) \to {\cal G}_{\zeta} (X;t;\mu)$.
At one loop,  the set of 
evolution equations 
for the flavor-singlet light-ray 
operators has the following form (cf. \cite{gdr,bb}):
\begin{equation}
 \mu \, \frac{d}{d \mu} \,  
{\cal O}_a(0,z)    =
\int_0^1  \int_0^{1}  
\sum_{b} A_{ab}(u,v ) {\cal O}_b( uz, \bar v z) \,
\theta (u+v \leq 1) \, du \, d v  \,  , 
\label{74} \end{equation}
where $a,b =  g,  Q$ and $\bar v \equiv 1-v$, $\bar u \equiv 1-u$. 
For  flavor-nonsinglet  distributions,
there is no mixing, and their  
 evolution 
is generated  by the $QQ$-kernel  alone.
To calculate the kernels,
we incorporated the approach \cite{bb} based on 
 the background-field method.
Below  we present  our results 
in the form similar to that used in refs.\cite{gluon,npd}:
\begin{eqnarray} 
& \displaystyle  
A_{QQ}(u,v ) = \frac{\alpha_s}{\pi} \, C_F \left 
(1 + \frac3{2}\,  \delta( u)\delta(v) + 
\left  \{ \delta(u) \biggl[ \frac{\bar v}{v} - \delta (v) \int_0^1
\frac{d \tilde v}{\tilde v} \biggr ] 
+ \{ u \leftrightarrow v \} \right \} 
 \right ) \, , \label{aqq}
\\
& \displaystyle  A_{gQ}(u,v ) = \frac{\alpha_s}{\pi} \, C_F \biggl 
( \delta( u)\delta(v) - 2 \biggr  ) \, ,
\\ 
& \displaystyle  A_{Qg}(u,v ) = \frac{\alpha_s}{\pi} \, N_f \left 
( 1- u - v  \right ) \, , \\ 
& \displaystyle  A_{gg}(u,v ) = \frac{\alpha_s}{\pi} \, N_c \biggl (
4(1  -u -v) + \frac{\beta_0}{2 N_c} \,
\delta(u)\delta(v) 
+ \left  \{ \delta(u) \biggl[ \frac{\bar v^2}{v} - \delta (v) \int_0^1
\frac{d \tilde v}{\tilde v} \biggr ] 
+ \{ u \leftrightarrow v \} \right \} 
 \biggr  )  \, .
\label{agg} \end{eqnarray}
Independently, these  kernels were calculated by 
Blumlein, Geyer and Robaschik \cite{bgr,bgr2}.
Their results agree with ours. 

\section{Evolution equations for nonforward distributions}

Inserting the light-ray evolution  equations (\ref{74}) 
 between chosen  hadronic states
and parametrizing  matrix elements by appropriate
distributions, 
one can get the ``old'' DGLAP \cite{gl,ap,d}
 and BL-type \cite{bl,ohrn,gro} evolution kernels 
as well as  calculate  the 
new  kernels   $ \Delta W_{\zeta}^{ab}(X,Z)$ governing the 
evolution 
 of  nonforward parton distributions:
\begin{equation}
 \mu \frac{d}{d\mu} \,  {\cal G}_{\zeta}^a(X;t;\mu) =
\int_0^1  \, \sum_b \,  \Delta W_{\zeta}^{ab}(X,Z) \, 
{\cal G}_{\zeta}^b( Z;t;\mu) \, d Z \,  .
\label{76} 
 \end{equation}

Extracting $\Delta W_{\zeta}^{ab}(X,Z)$ from the
light-ray kernels $A_{ab}(u,v)$, 
one should take into account  
the  extra $(pz)$ factor in the rhs 
of  the  gluon distribution definition, which 
under the   Fourier 
transformation  with respect to  $(pz)$
results in the  differentiation 
$\partial / \partial X$. 
Thus,  it is convenient to   introduce first the auxiliary kernels
$  \Delta M^{ab}_{\zeta}(X,Z)$  directly related to  the    
light-ray  kernels $A(u,v)$  by 
\begin{equation}
 \Delta M^{ab}_{\zeta}(X,Z) = \int_0^1  \int_0^1 A_{ab}(u,v) \, 
\delta(X- \bar u Z + v (Z- \zeta)) \,
\theta(u+v \leq 1)
\, du \, dv   \  .
\label{77} \end{equation}
 The  $ \Delta W$-kernels are obtained from  the $\Delta M$-kernels  using 
\begin{eqnarray}
 \Delta W^{gg}_{\zeta}(X,Z)= \Delta M^{gg}_{\zeta}(X,Z) \  \    ,   \  \ 
 \Delta W^{QQ}_{\zeta}(X,Z)= \Delta M^{QQ}_{\zeta}(X,Z), \label{78} \\ 
\frac{\partial}{\partial X} \Delta  W^{gQ}_{\zeta}(X,Z) =
-  \Delta  M^{gQ}_{\zeta}(\widetilde X,Z)
\, d \widetilde X \ \  , \  \
 \Delta W^{Qg}_{\zeta}(X,Z)= - \frac{\partial}{\partial X} 
\, \Delta  M^{Qg}_{\zeta}(X,Z) \,  . 
\label{79} \end{eqnarray}
Hence, to get $ \Delta W^{gQ}_{\zeta}(X,Z)$ we should integrate 
$ \Delta M^{gQ}_{\zeta}(X,Z)$ with respect to $X$.
We fix the  integration constant by  the requirement
that $ \Delta W^{gQ}_{\zeta}(X,Z)$ vanishes  for $X>1$.
Then 
\begin{equation}
 \Delta W^{gQ}_{\zeta}(X,Z)= \int_X^1  \Delta M^{gQ}_{\zeta}(\widetilde X,Z)
\, d \widetilde X  \,  . \label{79A}
\end{equation}

Integrating  the 
 delta-function in eq.(\ref{77}) 
produces   four different types of the $\theta$-functions,
each of which 
corresponds to a specific component of the kernel governing 
the 
evolution of the nonforward 
distributions.

\section{ BL-type  evolution kernels} 

When  $\zeta =1$,  
${\cal G}_{\zeta}(X)$ reduces to a distribution amplitude
whose  evolution is governed by the 
BL-type  kernels:
\begin{equation}
 \Delta W_{\zeta =1}^{ab}(X,Z)=  V^{ab}(X,Z). 
\label{81} \end{equation}
Taking $\zeta =1$ in Eq.(\ref{77})  we obtain 
\begin{equation}
\Delta M^{ab}_{\zeta =1}(X,Z) \equiv U^{ab}(X,Z) 
= \int_0^1  \int_0^1 A_{ab}(u,v) \, 
\delta(X- \bar u Z - v (1-Z)) \,
\theta(u+v \leq 1)
\, du \, dv   \, . 
\label{82} \end{equation}

In fact,  the BL-type kernels  
appear as a  part of the  nonforward  kernel $W_{\zeta }^{ab}(X,Z)$
even in the general $\zeta \neq 1,0$ case.
As explained earlier, if   $X$  is  in the
region $X \leq \zeta$, 
then the  function ${\cal G}_{\zeta}(X)$ 
can  be treated as a distribution amplitude
$\Psi_{\zeta}(Y)$ with $Y= X/  \zeta$. 
For this reason, when both $X$ and $Z$ are smaller than $\zeta$,
 the kernels 
$W_{\zeta}^{ab}(X,Z)$ simply  reduce 
to the  BL-type  evolution kernels $V^{ab}(X/\zeta,Z/\zeta)$.
Indeed, the relation (\ref{77}) can be written as  
\begin{equation}
\Delta M^{ab}_{\zeta}(X,Z) = \frac1{\zeta} \int_0^1  \int_0^1 A_{ab}(u,v) \, 
\delta \left ({X}/{\zeta}- \bar u {Z}/
{\zeta}  - v ( 1-{Z}/{\zeta}) \right ) \,
\theta(u+v \leq 1)
\, du \, dv  \, . 
\label{85} \end{equation}
Comparing this expression with the representation 
for the $U^{ab}(X,Z)$ kernels, we conclude   that   
in the region where $X/\zeta \leq 1$ and $Z/\zeta \leq 1$,
the kernels $\Delta M^{ab}_{\zeta}(X,Z)$ are given by 
\begin{equation}
\Delta M_{\zeta }^{ab}(X,Z) |_{0 \leq \{X,Z \} \leq \zeta}  =
\frac1{\zeta} \, U^{ab} \left ({X}/{\zeta}, 
{Z}/{\zeta} \right )\, .
\label{86} \end{equation}

Now, using  the expressions  connecting the $\Delta W$-  
and $\Delta M$-kernels, 
we obtain the following  relations between the nonforward  evolution kernels
$\Delta W_{\zeta }^{ab}(X,Z)$ 
in the region $0 \leq \{X,Z\}  \leq \zeta$ 
and the BL-type  kernels $V^{ab}(X,Z)$:
\begin{eqnarray}
 \Delta W_{\zeta }^{QQ}(X,Z) = \frac1{\zeta} \,  V^{QQ}\left ({X}/{\zeta}, 
{Z}/{\zeta} \right )  \ ;  \ 
\Delta W_{\zeta }^{gQ}(X,Z) =   V^{gQ}\left ({X}/{\zeta}, 
{Z}/{\zeta} \right )  \ ;  \nonumber \\ 
\Delta W_{\zeta }^{Qg}(X,Z) = \frac1{\zeta^2} \,  V^{Qg}\left ({X}/{\zeta}, 
{Z}/{\zeta} \right )  \ ;  \ 
\Delta W_{\zeta }^{gg}(X,Z) = \frac1{\zeta} \,  V^{gg}\left ({X}/{\zeta}, 
{Z}/{\zeta} \right ) \, . 
\label{87} \end{eqnarray}

The kernels $V^{ab}(X,Z)$, in their turn, are  
derived  from the auxiliary  kernels $U^{ab}(X,Z)$. 
Due to the symmetry property  $A_{ab}(u,v) =  A_{ab}(v,u)$
 the kernels  $U^{ab}(X,Z)$ 
satisfy  $U^{ab} (\bar X, \bar Z) = U^{ab}(X,Z)$.
Hence,  it is sufficient to know 
the $U$-kernels in the $X \leq Z$ region only:
$$
U^{ab}(X,Z) = \theta (X \leq Z) \,  U_0^{ab}(X,Z)
+ \theta (Z \leq X) U_0^{ab}(\bar X, \bar Z) \, , 
$$
with the basic function
$ U_0^{ab}(X,Z) \equiv \theta (X \leq Z) \, 
 U^{ab}(X,Z)$  given by 
\begin{equation}
U_0^{ab}(X,Z) 
= \frac1{Z} \int_0^{X}   \, 
A_{ab} \left  ( \bar v - (X-v)/Z  ,v \right ) dv \, . 
\label{83} \end{equation}
Using  Eqs.(\ref{aqq})-(\ref{agg}), the  $A \to U_0$ conversion formulas 
\begin{eqnarray}
\delta(u) \, \delta(v) \to \delta(Z-X) \ \ , \  \  1 \to \frac{X}{Z} 
\ \ , \  \ 
\delta(u) \, \frac{\bar v}{v} \to 0 \ \ , \  \ 
\delta(u) \, \left ( \frac{\bar v}{v} \right )^2 \to 0 , \nonumber  \\
\delta(v) \,  \frac{\bar u}{u} \to \left 
(\frac{X}{Z}\right )  \frac1{Z-X}  \ \ , \  \ 
\delta(v) \,  \frac{\bar u^2}{u} \to \left 
(\frac{X}{Z} \right )^2  \frac1{Z-X}  \ \ , \  \ 
u+v \to \frac{X}{Z} \left ( 1 -  \frac{X}{2Z}  \right )   
\label{84} \end{eqnarray} 
and Eqs.(\ref{77})-(\ref{81}), (\ref{87}) we   get 
the BL-type  kernels 
\begin{eqnarray} 
&& \lefteqn{
V^{QQ}(X,Z) = \frac{\alpha_s}{\pi} \, 
C_F \, \left \{ \, \left [ \frac{X}{Z} 
\left ( 1 + \frac{1}{Z-X} \right )\, 
\theta\, (X <Z ) \right ]_+ \, + \, \{ X \to \bar X, Z \to \bar Z \} 
 \right \}  \, } \,
 , \label{88} \\ &&
V^{Qg}(X,Z) = \frac{\alpha_s}{\pi} \, N_f \, 
 \left \{ \, - \frac{X}{Z^2}  \, \theta\, (X <Z ) \, +
\frac{ \bar X}{\bar Z^2} \, \theta\, (X > Z )  
  \right \} \,  ,\label{89} \\ &&
V^{gQ}(X,Z) = \frac{\alpha_s}{\pi} \, C_F \, 
\left \{   \frac{X^2}{Z}  \, \theta\, (X <Z ) \, 
 - \,\frac{\bar X^2}{\bar Z} \, \theta\, (X > Z )  
  \right \}   , \label{90} \\ &&
V^{gg}(X,Z) = \frac{\alpha_s}{\pi} \,  N_c \, 
\biggl \{ 
 \frac{2 X^2 -X - Z}{Z^2} \, \theta\, (X <Z ) \,  + \, 
\left [ \frac{ \theta\, (X <Z )}{Z-X} \right ]_+ \,  
 + \, \{ X \to \bar X, Z \to \bar Z \}  \nonumber \\ && \hspace{4cm} + \ 
  \frac{\beta_0}{2N_c}  \, \delta(X-Z) \, 
\biggr \} \, ,  
\label{91} 
\end{eqnarray}
  calculated originally
in \cite{ohrn,gro} for flavor-singlet pseudoscalar 
meson distribution amplitudes.
With respect to integration over $0 \leq X \leq 1$,
the ``plus''-prescription for a function  $V(X,Z)$ is
defined by
\begin{equation}
[V(X,Z)]_+ = V(X,Z) \, -  \, \delta \, (X-Z) \int_0^1
V(Y,Z) \, dY \, .
\end{equation}

The BL-type kernels  also 
govern the evolution  in the region corresponding to transitions
from a fraction $Z$ which is larger
than $\zeta$ to a fraction $X$ which is smaller 
than $\zeta$. Indeed,  
using   the $\delta$-function to calculate the integral
over $u$, we get
\begin{equation}
\Delta M_{\zeta }^{ab}(X,Z) |_{X \leq \zeta \leq Z }
= \frac1{Z} \int_0^{X/\zeta}   \, 
A_{ab} \biggl  ( [1- X/Z -v(1-\zeta/Z)] \, ,v \biggr ) dv \, , 
\label{92} \end{equation}
which has the same analytic form (\ref{83}) 
as the expression for $M_{\zeta }^{ab}(X,Z) $
in the region $X \leq Z \leq \zeta$.
For $QQ, gg$ and $Qg$ kernels, this automatically means that
 $\Delta W_{\zeta }^{ab}(X,Z) |_{X \leq \zeta \leq Z }$ is given by the 
same  analytic expression as $\Delta W_{\zeta }^{ab}(X,Z) $ for $X<Z< \zeta$.
Because of integration in Eq.(\ref{79A}), 
to get  $\Delta W_{\zeta }^{gQ}(X,Z)$  one should also know 
$\Delta M_{\zeta }^{gQ}(X,Z) $ in the region $\zeta \leq X \leq Z$. 
However, our explicit calculation confirms that
 $\Delta W_{\zeta }^{gQ}(X,Z)$ in the transition region $X \leq \zeta \leq Z$ 
is given by the same expression as $\Delta W_{\zeta }^{gQ}(X,Z)$ for $X<Z \leq \zeta $.

In application to parton distributions related to nonforward 
matrix elements, X. Ji was the first \cite{ji2} who calculated analogous kernels
$P^{\prime} (x, \xi)$  which govern the 
evolution of his  off-forward  parton distributions $\tilde H(x,t;\mu)$ 
in the  $-\xi/2 < x < \xi/2$ region 
(in our  variables this region corresponds to  $0<X < \zeta$).
He used a direct momentum-representation  approach
in the light-cone gauge.  After proper redefinitions
(discussed in  \cite{npd}), we reproduced 
his expressions for the  first  three kernels.
For the gluon-gluon kernel,
our result formally differs from that obtained  by X. Ji \cite{ji2}.
However, due to the symmetry properties 
of the gluon distribution in the X. Ji approach,
the relevant integral vanishes and 
 the difference does not contribute 
to  the evolution. 
Blumlein {\it et al.} \cite{bgr}   derive the 
``extended'' BL-kernels \cite{drm} 
 from the light-ray 
evolution equations.
  For $X \neq Z$,  we  agree with their results
except for the $gQ$-kernel  and up to obvious misprints
in the $QQ$ and $gg$-kernels 
\footnote{We are grateful to 
J. Blumlein who informed us 
that the authors of 
ref.\cite{bgr} agree with our results.}.

\section{  Generalized  DGLAP kernels}

   When  $X > \zeta$,  
we can treat the asymmetric distribution
 function ${\cal G}_{\zeta}^a  (X)$ 
as  a generalization of the 
usual  distribution function $\Delta f_a(X)$ for a skewed  
 kinematics. 
Hence,   evolution in the region 
$\zeta < X \leq 1$,  
 $\zeta < Z \leq 1$ is close 
to that generated by the  DGLAP equation. 
In particular,  it has the basic property that
  the evolved fraction
$X$ cannot be larger than the original
fraction $Z$. The relevant kernels 
are given by 
\begin{equation} 
\displaystyle  
\Delta M_{\zeta }^{ab}(X,Z) |_{\zeta \leq X \leq Z \leq 1}
=   \frac{Z-X}{ZZ'} \int_0^{1}    \, 
A_{ab} \left   ( \bar w \, (1- X/Z)  \, , \, 
w\, (1- X'/Z')  \right ) dw \, , 
\label{94} \end{equation}
where $X' \equiv X - \zeta$ and $Z' \equiv Z - \zeta$
are the ``returning'' partners of the original
fractions $X,Z$. Note, that since  $Z-X = Z' -X'$,  the kernels
$\Delta M_{\zeta }^{ab}(X,Z)$ are given by functions symmetric 
with respect to the interchange  of $X,Z$ with  $X',Z'$.  
Using the table for transition from the $A_{ab}$-kernels
to the $\Delta M^{ab}$-kernels in the region $\zeta \leq X \leq Z \leq 1$
\begin{eqnarray} 
\delta(u) \, \delta(v)  \to \delta(Z-X)  \ \ ; \  \
1 \to \frac{Z-X}{ZZ'}   \ \ ; \  \
(u+v) \to \frac{Z-X}{2ZZ'} \left [2-  \frac{X}{Z} - \frac{X'}{Z'} \right ] 
 \ \ ; \  \ \nonumber  \\ 
\left (\delta(u) \,  \frac{\bar v}{v} + \delta(v) \, \frac{\bar u}{u}  \right ) 
\to \frac1{Z-X} \left [ \frac{X}{Z}+ \frac{X'}{Z'} \right ]  \, ;  
\left (\delta(u) \,  \frac{\bar v^2}{v} + \delta(v) \, \frac{\bar u^2}{u}  \right ) 
\to \frac1{Z-X} \left [ \left (\frac{X}{Z} \right )^2
+ \left (\frac{X'}{Z'} \right )^2 \right ] \, ,
\label{95} \end{eqnarray} 
and Eqs.(\ref{78}), (\ref{79}), we  
obtain   the 
kernels
$\Delta P^{ab}_{\zeta}(X,Z)\equiv \Delta W_{\zeta }^{ab}(X,Z) 
|_{\zeta \leq X \leq Z \leq 1}$:
\begin{eqnarray} 
&& \lefteqn{
\Delta P^{QQ}_{\zeta}(X,Z) = \frac{\alpha_s}{\pi} \, 
C_F \, \left \{ \,  \frac{1}{Z-X} 
\left [ 1 + \frac{XX'}{ZZ'} \right ]\, \theta \, (X<Z) 
 \right. } \nonumber \\ && \left. \hspace{4cm} 
+ \, \delta(X-Z) \left [\, \frac32 - 
\int_0^1  \frac{du}{u} - 
\int_0^1  \frac{dv}{v} \right ] \right \}  \rightarrow 
\frac1{Z} \Delta P_{QQ}(X/Z) \, , \label{96} \\ &&
\Delta P^{Qg}_{\zeta}(X,Z) = \frac{\alpha_s}{\pi} \, N_f \, \frac{1}{ZZ'}
 \left \{ \frac{X}{Z} + \frac{X'}{Z'}
 - 1 \right \} \, \rightarrow 
\frac1{Z^2}\Delta  P_{Qg}(X/Z) \, , \label{97} \\ &&
\Delta P^{gQ}_{\zeta}(X,Z) = \frac{\alpha_s}{\pi} \, C_F \, 
\left \{ \frac{X}{Z} + \frac{X'}{Z'} - \frac{XX'}{ZZ'}
  \right \} \, \rightarrow 
\frac{X}{Z}\Delta  P_{gQ}(X/Z) \,  , \label{98} \\ &&
\Delta P^{gg}_{\zeta}(X,Z) = \frac{\alpha_s}{\pi} \,  N_c \, 
\left \{ \left ( 2\left [ \frac{X}{Z} + \frac{X'}{Z'} \right ]
 \frac{Z-X}{ZZ'} 
+\frac1{Z-X} \left [ \left (\frac{X}{Z} \right )^2 +
 \left ( \frac{X'}{Z'} \right )^2 \right ] \right ) 
\, \theta(X<Z) \right. \nonumber \\ && \left. 
\hspace{4cm} + \ 
\delta(X-Z) \left [ \frac{\beta_0}{2N_c} - 
\int_0^1  \frac{du}{u} - 
\int_0^1  \frac{dv}{v}
  \, \right ]
\right \} \, \rightarrow 
\frac{X}{Z^2}\Delta  P_{gg}(X/Z) \, .
\label{99} \end{eqnarray}
The formally divergent integrals over $u$ and $v$ 
provide here the usual ``plus''-type
regularization of the $1/(Z-X)$ singularities.
The prescription following 
from Eqs.(\ref{94}),(\ref{95}) is that  combining the $1/(Z-X)$
and $\delta(Z-X)$ terms into
$[{\cal G}_{\zeta}(Z)-{\cal G}_{\zeta}(X)]/(Z-X)$  in 
 the convolution of 
$\Delta P_{\zeta}(X,Z)$ with ${\cal G}_{\zeta}(Z)$ 
one should change $u \to 1-X/Z$ and $v \to 1-X'/Z'$.

As expected, the  $\Delta P^{ab}_{\zeta}(X,Z)$ 
kernels  have a symmetric form. The arrows indicate
how the nonforward   kernels $\Delta P^{ab}_{\zeta}(X,Z)$
 are related to the DGLAP kernels in the $\zeta=0$ limit
when $Z=Z'$ and $X=X'$.
Deriving these relations, one should take into account that
 the gluonic asymmetric distribution function 
${\cal G}_{\zeta}^g(X)$ reduces in the  $\zeta \to 0$ limit 
to $X \Delta g(X)$ rather than to $\Delta g(X)$. 

After the appropriate redefinitions, 
we managed to  reproduce from our results 
all four kernels $\Delta  P_{ab}(x,\xi)$ 
(relevant to the $x > \xi/2 $ region)
calculated by  X. Ji  \cite{ji2}.

Note, that in the region $Z > \zeta$ the evolved fraction 
$X$ is always smaller than $Z$. Furthermore, if $Z \leq \zeta$ then also $X\leq \zeta$,
$i.e.,$ distributions in the $X > \zeta$ regions are not affected by 
the distributions in the $X < \zeta$ regions.
Hence,  information  about 
the initial 
distribution in the $Z > \zeta$ region
is sufficient for  calculating  its evolution in this region.
This situation may  be contrasted with the evolution of distributions
in the $Z < \zeta$ regions: in that case one should know the 
nonforward parton distributions in the whole domain $0 <Z <1$.

\section{Conclusions} 

In this letter, we discussed  the calculation 
of the evolution kernels $\Delta W_{\zeta}(X,Z)$ 
for nonforward parton distributions
${\cal G}_{\zeta} (X,t)$ sensitive to
parton helicities. We presented the 
 evolution kernels  for the relevant 
light-ray operators  and demonstrated how 
one can obtain from them the components of the nonforward
kernels $\Delta W_{\zeta}(X,Z)$. 
Our results have a transparent relation 
with DGLAP and BL-type kernels and a compact form convenient 
for further practical applications such as numerical studies 
 of the evolution 
of nonforward distributions. 

\section{Acknowledgements}

We thank A. Afanasev, J. Blumlein, X. Ji, I. Musatov, G. Piller  and D. Robaschik 
for discussions and correspondence. 
This work was supported by the US Department of Energy under contract
DE-AC05-84ER40150.

\end{document}